\documentclass[11pt]{article}
\usepackage{orcidlink}
\usepackage[T1]{fontenc}
\usepackage{lmodern}
\usepackage{microtype}
\usepackage{amsmath,amssymb}
\usepackage[numbers]{natbib}
\usepackage{hyperref}
\usepackage{geometry}
\usepackage{booktabs}
\usepackage{orcidlink}
\usepackage{hyphenat}

\usepackage{tikz}
\usetikzlibrary{arrows.meta,positioning,calc,fit}

\hypersetup{
  colorlinks=true,
  linkcolor=blue,
  citecolor=blue,
  urlcolor=blue
}

\title{Audited Skill-Graph Self-Improvement for Agentic LLMs via Verifiable Rewards, Experience Synthesis, and Continual Memory}
\author{
Ken Huang~\orcidlink{0009-0004-6502-3673} \\
\textit{DistributedApps.ai, OWASP} \\
Fairfax, VA, USA \\
\texttt{ken.huang@owasp.org}
\and
Jerry Huang~\orcidlink{0009-0005-6224-4785} \\
\textit{Kleiner Perkins} \\
\texttt{jerryhuang2k@gmail.com}
}
\date{}

\tikzset{
  asgibox/.style={
    draw,
    rounded corners=3pt,
    align=center,
    minimum height=7.5mm,
    inner xsep=10pt,
    inner ysep=6pt
  },
  asgiarrow/.style={-Latex, line width=0.9pt},
  asgidbl/.style={<->, line width=0.9pt},
  asgidiag/.style={-Latex, line width=0.9pt}
}

\begin{document}
\maketitle

\begin{abstract}
Reinforcement learning is increasingly used to turn large language models into interactive agents that act over long horizons, call tools, and manage memory under partial observability. Recent surveys frame this trend as a shift from single-turn generation toward sequential decision-making, and organize agentic reinforcement learning around capabilities such as planning, reasoning, tool use, self-improvement, and memory \citep{zhang2025landscape}. At the same time, systems work highlights that stable scaling of multi-turn, multi-task training requires asynchronous generation--training pipelines and unified interfaces for heterogeneous environments \citep{zhang2025agentrl,luo2025agentlightning}. Despite these advances, deployed self-improvement loops face an operational security problem: optimization pressure, distribution shift, and imperfect observability create incentives for reward hacking, brittle specialization, and untraceable behavioral drift that is difficult to reproduce, audit, or govern. This paper proposes \emph{Audited Skill-Graph Self-Improvement} (ASG-SI), an approach that treats self-improvement as iterative compilation of an agent into a growing directed skill graph whose promotions are gated by verifier-backed evidence. ASG-SI combines verifiable, decomposed rewards for tool-use correctness, outcome validity, skill reuse, and composition integrity; experience synthesis for scalable stress testing and curriculum coverage; and continual memory control to prevent unbounded context growth while retaining long-horizon credit assignment. The central operational mechanism is a verifier--auditor that replays candidate skills and compositions, produces minimally sufficient evidence bundles for promotion decisions, and enables reconstruction of shaped rewards from replayable artifacts rather than opaque preference signals.
\end{abstract}

\section{Introduction}
Agentic reinforcement learning reframes optimization from single-turn text generation into temporally extended decision-making where a policy must interpret observations, select actions such as tool calls and memory operations, and pursue delayed rewards. A key theme in recent overviews is that agent capabilities often treated as prompting tricks---planning, tool integration, memory, reflection, and reasoning---can be formalized as parts of an interactive policy and improved via reinforcement learning when rewards are available \citep{zhang2025landscape}. Parallel work emphasizes that real progress depends on infrastructure: multi-turn rollouts introduce complex control flow, heterogeneous environments, and throughput constraints, motivating fully asynchronous pipelines, standardized data interfaces, and decoupled execution and training \citep{zhang2025agentrl,luo2025agentlightning}. These trends raise a practical question for security-critical settings: when an agent is allowed to improve while deployed, what is the unit of improvement, and how can that unit be verified, reproduced, and governed under optimization pressure?

\begin{figure}[t]
\centering
\textbf{Figure \ref{fig:architecture} (page \pageref{fig:architecture}): ASG-SI system architecture.}
\begin{tikzpicture}[node distance=15mm]
  \node[asgibox] (task) {Task Stream};
  \node[asgibox, right=22mm of task] (policy) {Policy Agent};
  \node[asgibox, right=22mm of policy] (tool) {Tool APIs};

  \node[asgibox, below=18mm of policy] (mem) {Memory Store\\Skill Compiler};
  \node[asgibox, right=12mm of mem] (ver) {Verifier \& Auditor};
  \node[asgibox, right=22mm of ver] (graph) {Audited \\Skill Graph};

  \draw[asgiarrow] (task) -- (policy);

  \draw[asgidbl] (policy) -- (tool);

  \draw[asgidbl] (policy) -- node[right, fill=white, inner sep=1pt]{trajectories} (mem);

  \draw[asgiarrow] (mem) -- (ver);
  \draw[asgiarrow] (ver) -- node[above, fill=white, inner sep=1pt]{verified skills} (graph);

  \draw[asgidiag] (graph.west) -- node[above, sloped, fill=white, inner sep=1pt]{reuse} (policy.south east);

\end{tikzpicture}
\caption{ASG-SI system architecture. Online interaction is separated from offline skill compilation and verification. Improvements enter the system only through audited skill promotion.}
\label{fig:architecture}
\end{figure}
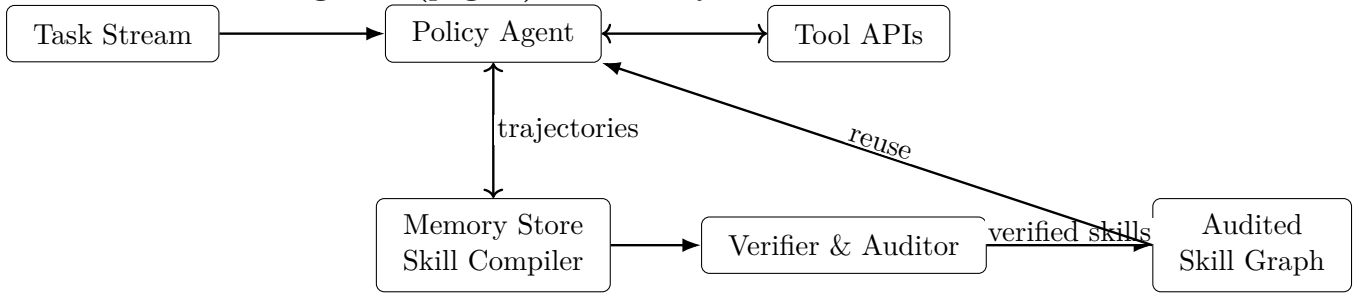

ASG-SI addresses this question by making improvement explicit and inspectable. Instead of treating self-improvement as only parameter drift, ASG-SI treats improvement as the accumulation of reusable skills with explicit interfaces, verifiable contracts, and auditable provenance. The proposal is motivated by an operational reality that appears across recent tool-learning and verifiable-reward lines of work: when rewards can be made reconstructible from artifacts such as tool transcripts, schema checks, or deterministic evaluators, learning can become more reliable and less dependent on subjective judgment, but the resulting systems still require strong measurement and contamination controls \citep{qian2025toolrl,wen2025rlvrreasoning,tu2025rlvrhidden}. ASG-SI operationalizes this insight by promoting new capabilities only after a verifier produces replayable evidence, and by structuring rewards so the same evidence can be used to reconstruct reward components. Figure~\ref{fig:architecture} provides a compact architecture that separates online interaction from offline compilation and verification.

\section{System Overview}
ASG-SI organizes an agentic system into four subsystems connected by logs and registries: a policy runtime that interacts with tasks and tools; a memory subsystem that supports bounded long-horizon operation; a skill compiler that extracts reusable skills from successful trajectories and normalizes them into canonical representations; and a verifier--auditor that replays candidates under controlled harnesses and emits evidence bundles. The audited skill graph is a directed multigraph whose nodes are skills with explicit interfaces and implementations, while edges encode composition constraints, dependency ordering, and guarded fallbacks. The key architectural goal is to ensure that the pathway from ``agent behavior'' to ``promoted improvement'' is mediated by artifacts that can be replayed and checked independently. That separation is aligned with the broader systems perspective in AgentRL and Agent Lightning, where execution and training are deliberately decoupled and unified via well-defined data interfaces \citep{zhang2025agentrl,luo2025agentlightning}. In ASG-SI, the verifier--auditor plays the role of an enforcement boundary: it does not rely on the policy model’s internal state, but on artifacts that can be reconstructed from logged trajectories, environment checks, and deterministic validators.

\section{Threat Model and Security Goals}
ASG-SI targets deployed agentic systems that self-improve while interacting with tools, external state, and memory. The baseline adversary model assumes that an attacker can influence task inputs and the distribution of tasks, can attempt to induce optimization pressure toward constraint-violating behavior, and can supply malicious or confusing content through tool outputs or external resources. This captures common deployment risks where agents are evaluated by outcome metrics while operating under constraints, and where a system that maximizes a metric can develop instrumental strategies that violate requirements to achieve the measured outcome. Recent work proposes explicit evaluation of such behaviors in multi-step, persistent environments, motivating security goals that go beyond single-turn refusal behavior and focus on outcome-driven constraint violations \citep{li2025odcvbench}. ASG-SI additionally considers supply-chain and integrity risks specific to self-improvement pipelines: a compromised skill artifact, a modified verifier, or tampered logs can create the appearance of improvement while degrading safety and reliability.

The security goals of ASG-SI are defined at the level of promoted artifacts and their provenance rather than at the level of a single model checkpoint. The first goal is \emph{auditability}, meaning that promotion decisions and reward assignments are tied to evidence bundles that can be replayed under a specified verifier version and environment harness. The second goal is \emph{integrity of improvement}, meaning that a skill is promoted only when it satisfies explicit interface contracts and passes replay-based tests that are independent of the policy runtime that generated the candidate. The third goal is \emph{controlled generalization}, meaning that skill reuse and composition are allowed only through interfaces whose preconditions and postconditions are checked, making failure localization possible at the node and edge level of the skill graph. The fourth goal is \emph{measurement reliability}, meaning that the evaluation protocol includes checks for reward gaming and measurement artifacts that can overstate gains under verifiable reward regimes, a concern raised in recent discussions of RL with verifiable rewards and measurement gaps \citep{tu2025rlvrhidden}. These goals assume that standard operational isolation is applied so that the verifier and evidence store are not directly writable by the policy runtime, and that versioning is enforced for verifiers, tools, and test harnesses.

\section{Motivation and Failure Modes}
A deployed self-improving agent must solve a control problem under partial observability where rewards are often sparse, delayed, or entangled with intermediate validity requirements such as tool-call schemas, policy constraints, or compliance checks. If optimization targets only final outcomes, training becomes unstable because many rollouts provide little learning signal, and agents can discover shortcuts that exploit evaluator blind spots. Progressive reward shaping formalizes a practical response by providing stage-wise objectives that first teach structural correctness of tool use and formatting, then correctness and quality, and only later efficiency, thereby reducing early-stage collapse in sparse-reward regimes \citep{zhuang2025prs}. Tool-learning work similarly emphasizes that reward design must reflect the structure of tool selection and parameterization, because coarse outcome matching can be too weak to guide intermediate actions \citep{qian2025toolrl}. ASG-SI follows this direction but extends it to operational governance by tying each reward component to replayable artifacts that also support promotion gating.

A second failure mode is forgetting under continual task streams. Continual learning benchmarks for agents highlight that evaluation must track performance over time as tasks evolve, and that catastrophic forgetting can manifest even when static benchmark scores appear stable \citep{joshi2025swebenchcl}. In deployed settings, forgetting can appear as regression in previously stable tool-use patterns, shifts in action selection under small context perturbations, or drift in memory contents that breaks long-horizon dependencies. Recent work trains memory operations as part of the policy using reinforcement learning, aiming to maintain performance while keeping memory bounded or structured \citep{zhou2025mem1,yan2025memoryr1}. ASG-SI incorporates this premise by treating memory operations as part of the auditable control loop and by making retention measurable through periodic replay testing of promoted skills and compositions.

A third failure mode is non-modular improvement. Even when an agent improves on an aggregate metric, it may be difficult to attribute which behaviors improved, whether the change generalizes, and whether the improvement can be reused safely. Skill library approaches attempt to make improvement modular by explicitly generating and reusing skills, and recent RL-based formulations integrate skill generation and utilization into the training objective \citep{wang2025sage}. ASG-SI extends modularity with verification and auditing so that skills are not only added but promoted through evidence-backed checks, and so that the operational unit of improvement is a verified skill artifact with an explicit interface and provenance trail rather than an implicit shift in model behavior.

\section{Audited Skill-Graph Self-Improvement}
ASG-SI treats each iteration as a compilation-and-verification cycle in which the policy runtime interacts with tasks using a combination of direct reasoning, tool calls, and memory operations while logging full trajectories that include tool transcripts, intermediate artifacts, and memory updates. The skill compiler extracts candidate skills by identifying reusable subsequences, normalizing them into canonical programs or templates, and assigning explicit interfaces with preconditions and postconditions. These interfaces are designed to be checkable by the verifier, including schema constraints for tool calls and structural invariants for intermediate artifacts. The verifier--auditor then evaluates candidates on held-out tasks, contract checks, and controlled perturbations, producing an evidence bundle that includes deterministic test results when applicable, tool schema validation logs, and cryptographic hashes of artifacts or transcripts. Only candidates that pass promotion criteria are added to the audited skill graph, which becomes an explicit runtime substrate for future reuse.

\begin{figure}[t]
\centering
\textbf{Figure \ref{fig:loop} (page \pageref{fig:loop}): Iterative self-improvement loop in ASG-SI.}
\begin{tikzpicture}[node distance=16mm]
  \node[asgibox] (env) {Environment};
  \node[asgibox, right=20mm of env] (agent) {Policy Agent};
  \node[asgibox, below=14mm of agent] (log) {Trajectory Log};
  \node[asgibox, below=14mm of log] (comp) {Skill Compiler};
  \node[asgibox, below=14mm of comp] (ver) {Verifier \& Auditor};
  \node[asgibox, right=22mm of ver] (graph) {Skill Graph};
  \node[asgibox, right=22mm of comp] (train) {RL Trainer};

  \draw[asgidbl] (agent) -- (env);
  \draw[asgiarrow] (agent) -- (log);
  \draw[asgiarrow] (log) -- (comp);
  \draw[asgiarrow] (comp) -- (ver);
  \draw[asgiarrow] (ver) -- node[above, fill=white, inner sep=1pt]{pass} (graph);
  \draw[asgiarrow] (ver) -- (train);
  \draw[asgiarrow] (train) -- (agent);

  \draw[asgidiag] (graph.north west) -- node[above, sloped, fill=white, inner sep=1pt]{reuse} (agent.south);

\end{tikzpicture}
\caption{Iterative self-improvement loop in ASG-SI. Candidate skills are compiled from trajectories, verified by replay and contract checks, and then promoted. Evidence bundles drive shaped rewards and support reproducibility.}
\label{fig:loop}
\end{figure}
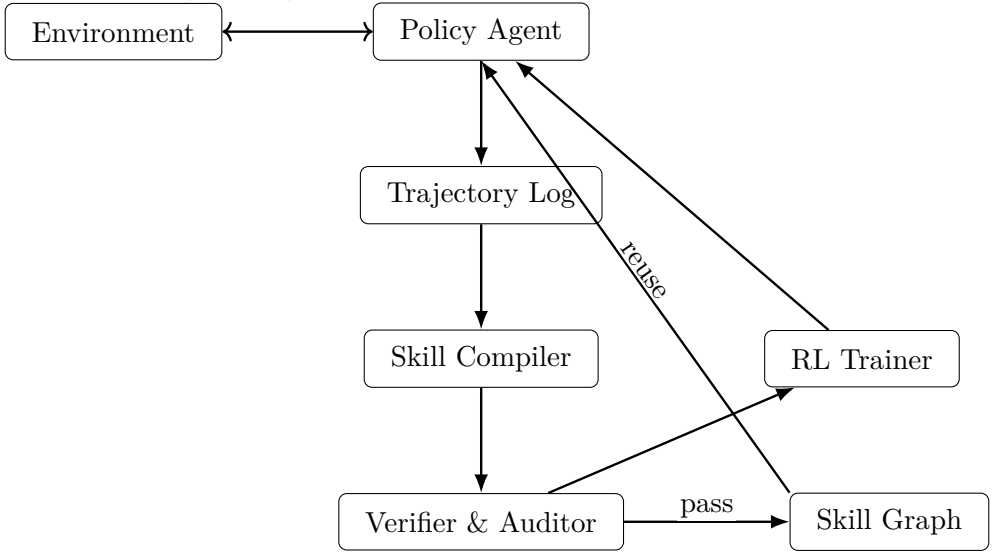

The emphasis on verification before promotion is motivated by the operational gap between ``learning improves a metric'' and ``learning produces a reusable, governable capability.'' In ASG-SI, a promoted skill is a callable unit whose interface and evidence can be inspected and whose execution can be replayed under a specified harness. This design also enables decomposition of success attribution: the system can distinguish success attributable to direct reasoning from success attributable to verified skill reuse and verified composition. Figure~\ref{fig:loop} depicts the iterative loop in which verification produces both promotion decisions and the evidence used for reward shaping.

\section{Verifier-Backed Reward Construction}
ASG-SI uses a decomposed reward designed to be reconstructible from evidence bundles. The tool validity component scores whether tool calls satisfy schemas, whether arguments are well-typed, and whether outputs are used consistently across subsequent steps; this follows the observation that structured rewards for tool use are often needed because final-answer matching can be too coarse to guide policy improvement \citep{qian2025toolrl}. The outcome verification component scores task success using deterministic evaluators when available, such as unit tests, explicit environment checks, or exact-match scoring in structured tasks. The skill reuse component rewards invoking a verified skill under its preconditions and producing outputs that satisfy postconditions, turning reuse into a contract-respecting action rather than a heuristic. The composition component rewards multi-skill chains that satisfy interface contracts at each edge and penalizes contract violations that would otherwise be hidden behind end-to-end success. The memory discipline component penalizes unbounded context growth and rewards memory operations that preserve success under bounded memory, aligning with approaches that treat memory updates and retrieval decisions as learnable actions under RL objectives \citep{zhou2025mem1,yan2025memoryr1}.

Progressive shaping schedules these components across training phases, prioritizing structural validity early and shifting weight toward correctness, composition integrity, and efficiency later. This aligns with progressive reward shaping methods in agentic RL that provide dense, stage-wise feedback to overcome sparse binary rewards and stabilize optimization \citep{zhuang2025prs}. The verifier--auditor is responsible for emitting the artifacts needed for reconstruction: schema-validation logs, test outcomes, interface checks, and hashes of intermediate artifacts. This design also supports stronger measurement discipline, because the same artifacts can be used to detect inconsistencies between claimed success and verifiable success and to audit whether reward components were computed as specified, a point that becomes important when verifiable reward regimes are evaluated under parity-controlled protocols \citep{tu2025rlvrhidden}.

\section{Skill Graph Representation}
\begin{figure}[t]
\centering
\textbf{Figure \ref{fig:skillgraph} (page \pageref{fig:skillgraph}): Skill graph composition example.}
\begin{tikzpicture}[node distance=22mm]
  \node[asgibox] (s1) {Skill A\\Localization};
  \node[asgibox, right=18mm of s1] (s2) {Skill B\\Synthesis};
  \node[asgibox, right=18mm of s2] (s3) {Skill C\\Test Runner};
  \node[asgibox, right=18mm of s3] (goal) {Task\\Success};

  \node[asgibox, below=14mm of s2] (fb) {Fallback Skill\\Minimal Fix};

  \draw[asgiarrow] (s1) -- node[above, fill=white, inner sep=1pt]{symbols} (s2);
  \draw[asgiarrow] (s2) -- node[above, fill=white, inner sep=1pt]{patch} (s3);
  \draw[asgiarrow] (s3) -- node[above, fill=white, inner sep=1pt]{pass} (goal);

  \draw[asgidiag] (s2) -- node[right, fill=white, inner sep=1pt]{guarded} (fb);

\end{tikzpicture}
\caption{Skill graph composition example. Nodes are verified skills with explicit interfaces; edges encode compositional contracts and guarded fallbacks.}
\label{fig:skillgraph}
\end{figure}
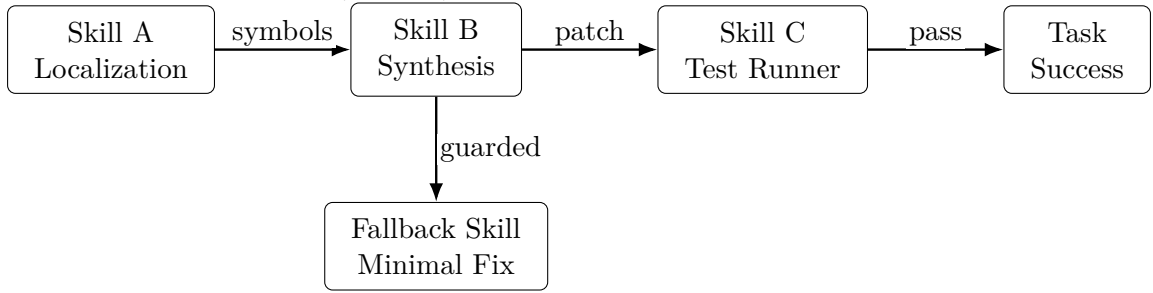

The audited skill graph provides a concrete substrate for modular improvement, failure localization, and governance. Nodes correspond to skills with explicit interfaces, canonical implementations, and verification reports; edges encode compatibility and composition constraints such as required intermediate types, ordering constraints, or guarded fallbacks. By explicitly representing interfaces, ASG-SI can verify skills and compositions by replay under a harness, localize regressions to specific nodes or edges, and measure improvement as the accumulation of verified reusable units rather than only the change in an aggregate benchmark score. This is a practical response to the non-modularity failure mode: when improvement is represented as a graph of artifacts with verifiable contracts, it becomes feasible to ask which new capability was added, which tasks it applies to, and what evidence supports its claimed behavior. Figure~\ref{fig:skillgraph} illustrates a common multi-step composition pattern with a guarded fallback, which is meant to capture realistic agent behavior where a primary strategy can fail and a conservative alternative is invoked under explicit guard conditions.

\section{Auditing and Evidence Bundles}
\begin{figure}[t]
\centering
\textbf{Figure \ref{fig:audit} (page \pageref{fig:audit}): Auditable improvement trace.}
\begin{tikzpicture}[node distance=16mm]
  \node[asgibox] (traj) {Trajectory};
  \node[asgibox, below=12mm of traj] (cand) {Candidate Skill};
  \node[asgibox, below=12mm of cand] (replay) {Verifier Replay};

  \node[asgibox, right=22mm of replay] (evid) {Evidence\\Bundle};
  \node[asgibox, below=12mm of replay] (dec) {Pass / Fail};
  \node[asgibox, right=22mm of dec] (rew) {Reward\\Components};
  \node[asgibox, below=12mm of dec] (graph) {Promote to\\Skill Graph};

  \draw[asgiarrow] (traj) -- (cand);
  \draw[asgiarrow] (cand) -- (replay);
  \draw[asgiarrow] (replay) -- (dec);
  \draw[asgiarrow] (replay) -- (evid);
  \draw[asgiarrow] (evid) -- (rew);
  \draw[asgiarrow] (dec) -- node[right, fill=white, inner sep=1pt]{pass} (graph);

  \node[asgibox, above=30mm of rew] (train) {RL Update};
  \draw[asgiarrow] (rew) -- (train);

\end{tikzpicture}
\caption{Auditable improvement trace. Promotion decisions and shaped rewards are derived from replayable evidence bundles rather than opaque preference signals.}
\label{fig:audit}
\end{figure}
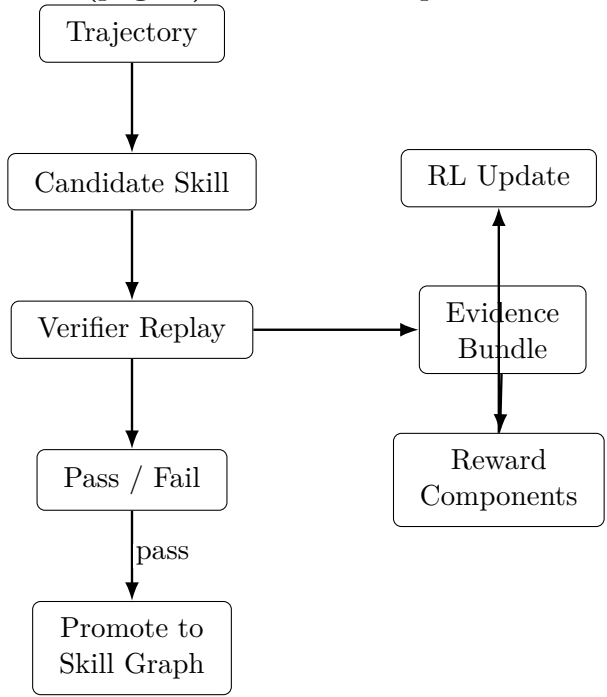

Auditing is treated as a first-class output that binds promotion decisions, reward construction, and reproducibility into a single trace. Each promotion decision is tied to a minimally sufficient evidence bundle that includes tool schemas and arguments, tool outputs (or output hashes when outputs are large), contract checks for skill interfaces, deterministic test results when applicable, and verifier identity with versioning. These evidence bundles allow reward reconstruction, enable third-party replay, and make regression detection operational because the same tests that justified promotion can be rerun periodically or on new model checkpoints. This design is motivated by the observation that outcome metrics can be gamed under optimization pressure and that robust evaluation often requires explicit measurement and provenance checks when verifiable rewards are used at scale \citep{tu2025rlvrhidden}. In settings where agents might be incentivized to violate constraints to achieve outcomes, explicitly recording constraint checks and incorporating them into reward and promotion gating provides a practical mechanism aligned with benchmarks that target outcome-driven constraint violations in multi-step environments \citep{li2025odcvbench}. Figure~\ref{fig:audit} shows the flow from trajectories to candidates to verifier replay, and then to promotion and shaped rewards derived from evidence.

\section{Evidence Integrity and Tamper-Evidence Design}
Evidence integrity is a security boundary for any audited self-improvement system because tampered evidence can create the appearance of verified improvement while undermining safety and reliability. ASG-SI treats each evidence bundle as a structured manifest that includes a canonical ordering of fields, cryptographic hashes of artifacts and transcripts, and explicit version identifiers for the verifier, tool schemas, and environment harness. This design supports tamper-evidence in two ways: it makes post-hoc modification detectable by recomputing hashes and comparing manifests, and it makes provenance comparable across runs by binding results to a specific verifier version. The integrity requirement becomes more salient as systems adopt asynchronous pipelines, because decoupling execution and training improves throughput but also increases the importance of trace integrity and explicit staleness and provenance controls in distributed rollouts \citep{zhang2025agentrl,luo2025agentlightning}. For security-category submissions, this section can be operationalized as concrete requirements on log append-only storage, key management for signing, and separation of duties between policy runtime and verifier, with the core claim remaining that evidence integrity is necessary to make auditability meaningful under adversarial pressure.

\section{Scaling with Experience Synthesis}
Experience synthesis provides a path to scale training and verification when real-environment interaction is expensive or rate-limited. DreamGym proposes synthesizing diverse agent experiences by distilling environment dynamics into an experience model and combining synthetic transitions with replay buffers that are periodically re-anchored with real interactions \citep{chen2025dreamgym}. In ASG-SI, synthesized experience is treated as complementary to real rollouts rather than a replacement, because the security objective is not only performance but also reliable evidence and bounded divergence between synthetic and real distributions. Synthesis is used primarily to stress-test skill interfaces and compositions, probe boundary conditions that are rare under natural task distributions, and expand coverage for adversarial and edge-case behaviors. Periodic re-grounding with real rollouts recalibrates verifier thresholds and reduces the risk that a skill is promoted based on synthetic artifacts that do not transfer to the real harness. This framing also supports a more security-aligned use of synthesis: it can generate adversarially structured tasks that exercise constraint boundaries and tool misuse patterns, and those tasks can be integrated into held-out verification suites used for promotion gating.

\section{Continual Learning and Retention}
Continual learning makes forgetting observable because evaluation tracks performance along a task stream rather than on a static test set. SWE-Bench-CL exemplifies this by organizing software engineering issues chronologically and measuring whether agents retain and transfer capabilities as repositories evolve \citep{joshi2025swebenchcl}. ASG-SI addresses retention through a dual mechanism. First, verified skills persist as callable artifacts in the graph, making them available as stable fallbacks even if the base policy’s behavior drifts. Second, memory operations are controlled so that long-horizon dependencies can be retained without unbounded context growth, aligning with approaches that optimize memory updates and retrieval through reinforcement learning rather than fixed heuristics \citep{zhou2025mem1,yan2025memoryr1}. This combination enables measurable retention: verified skills can be periodically replay-tested to detect regressions at the artifact level, while reuse statistics can quantify whether retained skills remain operationally relevant as the task stream shifts.

\section{Implementation Considerations}
ASG-SI is designed to integrate with asynchronous agentic RL infrastructures where execution and training are decoupled and connected by standardized data interfaces. AgentRL emphasizes scalable multi-turn, multi-task training with asynchronous generation--training pipelines and heterogeneous environment support \citep{zhang2025agentrl}, while Agent Lightning describes disaggregating agent execution from RL training and defining a unified data interface that enables integration with diverse agent runtimes \citep{luo2025agentlightning}. ASG-SI is compatible with this direction by treating the compiler and verifier as sidecar services that consume trajectory logs and emit skill artifacts, verifier reports, and evidence bundles. The policy runtime consults the audited skill graph for invocation candidates, while the trainer consumes evidence bundles to compute shaped rewards and to support parity-controlled evaluation where the same evidence can be reused to validate outcomes and intermediate constraints.

A reference implementation of the paper is available at \href{https://github.com/kenhuangus/ASG-SI}{https://github.com/kenhuangus/ASG-SI}. The repository includes a fully runnable minimal prototype (\texttt{asg\_si\_demo.py}) that mirrors the ASG-SI architecture in a single file and makes all intermediate artifacts explicit. The prototype defines a verifiable environment that generates tasks with deterministic ground truth and exact-match verification, which serves as a concrete instantiation of verifiable outcome reward. It implements a small tool registry with explicit schemas and pure-function tools to keep replay deterministic and to make schema-validation meaningful. It represents interaction as trajectories composed of typed steps (tool call, skill call, or direct action) and logs sufficient data to reconstruct both outcome correctness and intermediate validity. It implements a bounded memory component that retains a limited number of compact notes and exposes a summarized hint to the agent, providing a concrete baseline for controlled memory growth that can later be replaced by learned memory operations without changing the audit boundary.

The skill compiler in the prototype demonstrates the ``compile'' step as extraction of a canonical program from a successful trajectory. Specifically, when a task is solved correctly via a single tool call, the compiler constructs a program template that maps tool arguments to task input keys and assigns an explicit interface describing required inputs and output type. This is then passed to a verifier--auditor that replays the candidate skill on a held-out set of tasks of the same kind and produces a verification report including per-test outcomes, pass rate, check timestamps, and a stable hash of the program representation. Each verification event is written as a JSON audit trace to an append-only audit directory, making the evidence bundle concrete and inspectable. The reward shaping function in the prototype decomposes reward into structural validity, outcome correctness, reuse bonus, and a step-count penalty, and it logs each component so that reward reconstruction can be performed from evidence and trajectory logs. The prototype also includes an optional model-based policy that calls a local model endpoint and parses outputs into tool calls or direct answers, allowing the same audit and verification machinery to be exercised under realistic tool-use errors such as mis-specified arguments or wrong tool selection. The implementation is intentionally structured so that the verifier boundary does not depend on the model’s internal reasoning: promotion and reward components are determined by artifacts that can be replayed and validated, which aligns with the paper’s claim that audited improvement traces can support reproducible evaluation and governance.

\section{Evaluation Plan}
Evaluation for ASG-SI must measure capability growth, retention under continual streams, and constraint adherence under optimization pressure, while also measuring the integrity of the self-improvement pipeline itself. Capability growth should be decomposed into success attributable to direct reasoning, success attributable to verified skill reuse, and success attributable to verified composition, because the thesis is that durable improvement manifests as growth in auditable reusable artifacts rather than only aggregate score increases. Tool validity metrics should include schema-correctness rates, argument-type correctness, and tool-output utilization consistency, reflecting evidence-backed intermediate validity rather than only final outcomes \citep{qian2025toolrl}. Retention should be measured under a task stream such as SWE-Bench-CL by tracking performance as tasks evolve and by periodically replay-testing promoted skills on fixed verification suites to detect regressions, capturing both policy drift and artifact stability \citep{joshi2025swebenchcl}. Constraint adherence should be evaluated in environments designed to elicit outcome-driven violations where agents may be incentivized to bypass rules to maximize metrics, and evidence bundles should be audited for signatures of reward hacking such as inconsistent tool transcripts or contract violations masked by end outcomes \citep{li2025odcvbench}. Because recent analyses note that verifiable reward gains can be overstated without parity-controlled protocols and contamination checks, ASG-SI evaluation should report evidence-based reconstruction rates, verifier reproducibility across versions, and sensitivity to harness changes \citep{tu2025rlvrhidden}. A practical composite metric implied by the design is an audited improvement rate that combines the rate of newly promoted verified skills with their reuse rate on held-out tasks, which captures both creation of artifacts and their operational utility.

\section{Limitations and Open Failure Cases}
ASG-SI depends on the integrity and isolation of the verifier and the evidence store; if the verifier is compromised or if evidence can be rewritten by the policy runtime, auditability becomes a surface property without enforcement power. Replay-based verification introduces overhead, and nondeterministic environments require controlled harnesses, record-and-replay mechanisms, or robust statistical testing to keep evidence meaningful. Verifiable rewards can also create measurement incentives that shift behavior toward what is easily verified rather than what is desired, and recent work highlights that careful evaluation design is needed to avoid overstating gains or missing hidden costs under verifiable reward regimes \citep{tu2025rlvrhidden}. Evidence bundles may contain sensitive data in realistic deployments, requiring access control, retention policies, and careful minimization of what must be stored to support replay. Finally, skill interfaces can be under-specified: if preconditions and postconditions are too weak, a skill can pass tests yet fail in edge cases; if they are too strict, useful skills may never be promoted. These tradeoffs motivate future work on interface inference, adversarial verification suites, and principled composition typing for complex tool and environment interactions.

\section{Conclusion}
ASG-SI reframes agentic self-improvement as accumulation of verifiable, reusable capabilities rather than uncontrolled parameter drift. By compiling behavior into an audited skill graph, shaping rewards with verifier-backed evidence, scaling training with a grounded experience synthesis strategy, and controlling memory growth for long-horizon interaction, the approach aims to make self-improvement measurable, reproducible, and governable in security-relevant deployments.

\bibliographystyle{plainnat}
\bibliography{references}

\end{document}